\documentstyle[aps]{revtex}
\begin{document}

\title{Production of neutral and charged Higgs bosons of the 
       MSSM at the future $\bf e\gamma$ colliders 
       }

\author{U. Cotti$^{1,3}$,\,
        J.L. Diaz-Cruz$^2$\protect\thanks{e-mail: ldiaz@sirio.ifuap.buap.mx} 
and \,  J.J. Toscano$^4$ }

\address{$^1$SISSA-ISAS, via Beirut 2-4 I-34013, Trieste, Italy\\
         $^2$Instituto de F\'{\i}sica, BUAP, A.P. J-48, 72570 Puebla, 
         Pue. M\'exico \\
         $^3$Dpto. de F\'{\i}sica, CINVESTAV-IPN, M\'exico D.F.\\
         $^4$ Facultad de Ciencias F\'\i sico Matem\' aticas, BUAP, 
              A. P. 1152,Puebla, Pue., M\' exico}

\date{\today}

\maketitle

\begin{abstract}
 A complete study for the production of neutral ($h^0,H^0, A^0 (=\phi^0_i)$)
and charged Higgs ($H^\pm$)
bosons at electron-photon colliders is presented in
the context of the minimal supersymmetric standard model.
 A particular choice of the non-linear $R_{\xi}$-gauge is used to
evaluate the amplitudes of the reaction $e\gamma \to e \phi^0_i$.
 The resulting cross section indicates that it will be possible to detect
a signal from the neutral Higgs bosons for most regions
of parameter space at the future linear colliders with $\sqrt{s}=500$ GeV
through the reaction
$e\gamma \to e \phi^0_i$. This reaction also offers the interesting
possibility to measure the Higgs mass through
the detection of the outgoing electron.
 The production of the charged Higgs boson ($H^+$) through the
 reaction $e\gamma \to \nu_e H^\pm$ has in general
smaller values for the cross section, which seems more difficult to
observe. 

\end{abstract}

\section{Introduction}
 The search for Higgs bosons at future colliders has become 
the focus of extensive studies, because of its importance 
as a test of the mechanism of electroweak symmetry 
breaking~\cite{kane:90}.
 The detection of the full spectrum of scalars seems
necessary in order to determine the nature of the physics
that lies beyond the {\sl standard model} (SM). 
 Among the extensions of the SM, supersymmetry has received increasing 
attention, not only because of its aesthetical properties as a 
field theory, but also because the naturalness problem 
of the SM can be alleviated within the 
so-called {\sl minimal supersymmetric standard model} 
(MSSM)~\cite{Martin:1997ns}.

 The MSSM contains two Higgs doublets, whose physical spectrum 
includes a charged pair $(H^\pm)$, two neutral CP-even scalars 
($h^0$, $H^0$), and one pseudoscalar 
($A^0$).
 The MSSM Higgs sector is determined at tree-level by two 
parameters, which are nowadays chosen as the $A^0$ mass and
$\tan\beta$ (the ratio of the vevs. of the two Higgs doublets). 
 This in turn fixes the values of the neutral Higgs mixing angle 
$\alpha$ and the remaining Higgs masses, which obey the 
tree-level relations, $m_{h^0} \leq m_Z \leq m_{H^0}$, 
$m_{h^0}\leq m_{A^0} \leq m_{H^0}$, $m_W \leq m_{H^\pm}$.
 However, these relations are substantially modified when the 
effect of radiative corrections is 
included~\cite{Okada:1991vk,Haber:1991aw,Ellis:1991nz}.
 In particular, it makes possible that $m_Z\leq m_{h^0}$, 
$m_{A^0} \leq m_{h^0}$, and for some regions of parameters 
$m_{h^0}$ could even reach a value of about 
130~GeV~\cite{Kunszt:1992qe,Zhang:1998bm,Heinemeyer:1998jw}.

 Current LEP2 limits on the Higgs masses are of about 90~GeV 
for the light scalar $(h^0)$, and, depending on the value
of $\tan\beta$, up to about 85~GeV for 
$A^0$~\cite{PepeAltarelli:1999ii}. 
 However LEP2 will be able to cover the region up to
$m_{h^0} \lesssim 110$ GeV~\cite{Ellis:1999bk}. 
 Recently, it was found that
Tevatron can be used to test a significant portion of 
parameter space through the reaction 
$p \bar{p}\to Wh^0 + X$~\cite{Carena:1998gk},
moreover the possibility to perform b-tagging with
a high efficiency has opened the window to detect
the mode $p{\bar p} \to b{\bar b} h^0 + X$ at 
LHC~\cite{Dai:1996rn}
and it will allow
to test the large $\tan\beta$ region of parameters
at Tevatron~\cite{Balazs:1999nt}. 
 On the other hand, it has been shown~\cite{Spira:1997dg}
that LEP2 plus LHC results will be able to cover almost all the
MSSM Higgs sector parameter space, and through a combination
of the reactions
$ p \bar{p} \to t \bar{t} +h^0 (\to \gamma \gamma) + X $,
$ p \bar{p} \to h^0 (\to ZZ^*) + X $%
~\cite{Gunion:1992er,Diaz-Cruz:1993hc,Dittmar:1997ss}
and weak boson fusion~\cite{Plehn:1999nw},
the full region will be covered.
 At future linear colliders like NLC~\cite{Kuhlman:1996rc},
TESLA~\cite{Accomando:1997wt} or JLC~\cite{Tauchi:1998ty}, it 
will be possible to search for neutral and charged Higgs bosons, 
through the production reactions 
$e^+e^- \rightarrow h^0 Z, h^0 \gamma$
~\cite{Haber:1996qb,Barroso:1986et}
and $e^+e^- \rightarrow H^+ H^-$~\cite{Komamiya:1988rs}.

 The future linear colliders can also operate in the $e\gamma$ 
mode~\cite{Giacomelli:1993ip}; in this case the production
of the SM Higgs boson has been studied through the 
$\gamma\gamma \rightarrow \phi^0$ 
mechanism~\cite{Eboli:1993jt,Hernandez:1996br}, 
and also with the full 1-loop two-body reaction 
$e \gamma \rightarrow \phi^0_i e$~\cite{%
Cotti:1997er,%
Gabrielli:1997ix,%
Liao:1998ss}.
 The production of the pseudoscalar Higgs boson $A^0$
through the reaction $e \gamma \rightarrow A^0 e$
has been studied too~\cite{Dicus:1996in}, using the
photon pole approximation.
 Even though, these reactions can occur at tree-level,
they receive the main contributions at one-loop level where the
heavy particles of the model play an important role.  
 These processes could be used to measure the  
couplings:
$\phi^0\gamma^*\gamma$, $\phi^0Z^*\gamma$%
~\cite{Gabrielli:1997ix,Banin:1999ap,Gabrielli:1999qi},
which constitute important one-loop predictions of the theory, 
and are also sensitive to the effects of new physics.
 Moreover, because of the possibility to measure the electron
momenta, this reaction offers the possibility
to determine the Higgs mass with high precision.

 In this paper we present a detailed analysis of the production of the
neutral and charged Higgs of the MSSM  at $\gamma e^-$ colliders
through the reactions $e\gamma \to e \phi^0_i$
($\phi^0_i=h^0,H^0,A^0$) and
$e\gamma \to \nu H^\pm$, at the one-loop level.
Our goal is to determine the regions of parameter space where
a signal is detectable, and also to find out where it will be possible to
distinguish between the MSSM and SM Higgs signals.
 We shall assume that the super-partners are heavy,
and thus decouple from the amplitudes~\cite{Dobado:1998tr},
however the effect of squarks will be included
in the Higgs effective potential using 
the approximations presented in Ref.~\cite{Barbieri:1991tk}.

 The organization of the paper is as follows: Section 2 discusses
the production of neutral Higgs bosons.
 It includes a discussion of the region of parameter space that
can be excluded, and also on the determination of the
Higgs mass. 
 Section 3 is devoted to the production
of the charged Higgs, whereas the conclusions of our
work are presented in Section 4.
 Details about the non-linear gauge used
in the evaluation of the 1-loop amplitudes as well as 
the explicit formulae for the various amplitudes 
are presented in the Appendices.

\section{The neutral Higgses}
 We now proceed to present the results of our calculation of the amplitudes
of the reactions:
\begin{eqnarray}
 \gamma
  (k_1)+e^-(p_1) & \rightarrow & \phi^0(k_2)+e^-(p_2),
\end{eqnarray}
where $\phi^0$ denotes any of $h^0,H^0,A^0$, and we have also displayed
the notation for incoming and outgoing momenta.
 We have organized the calculation according to the 
$\rm U_{em}(1)$ gauge invariance, thus the diagrams are grouped as
follows:

\begin{enumerate}
 \item three-point diagrams characterized by the $\phi^0\gamma^*\gamma$
       and by the $\phi^0Z^*\gamma$ coupling (Fig. 1-a) ;
 \item Z and W-mediate box diagram and its related triangle graphs
       (Fig. 1-b,c,d),
\end{enumerate}

 The triangle graphs related to the $Z$- and $W$-mediated box diagrams are
the one-loop $\phi^0 e^*e$ 3-point functions.
 These groups of diagrams are finite and gauge invariant by themselves.
 The group (1) receives contributions from loops of charged fermions,
$W$ gauge boson, and charged Higgs boson $H^{\pm}$.
 On the other hand, the group (2) is sensitive to the $\phi^0
W^+W^-$ and $\phi^0 ZZ$ vertices, respectively.
 For $A^0$ there are contributions from fermionic triangles
only, thus only group (1) appear.

 In the linear gauge, the reaction $e\gamma \to e \phi^0_i$
receives also contributions coming from the reducible diagrams 
with the $Z^*\gamma$ self-energy, which is an extra complication because
one needs to perform a renormalization of this term. 
 However, in the nonlinear gauge this
term is absent as consequence of the explicit $\rm U_{em}(1)$ gauge
symmetry in the W, goldstone and ghost sectors.
 Notice that there are no contributions coming from the charged Higgs
boson to the group (2).
 This happens because we use the approximation $m_e=0$ and also because
there is no $\gamma W^{\pm}H^{\pm}$ vertex at tree-level.%
\footnote{We could think on possible contributions arising from reducible
diagrams that include the $\gamma^*\phi^0$ and $Z^*\phi^0$ self-energies
(with the virtual fields tied to the electronic line).
 However, it can be shown that these terms always vanish.}

\subsection{Production of $h^0,H^0$}

 The Mandelstam variables used in this calculation are defined by
$s=(k_1+p_1)^2$, $t=(k_1-k_2)^2$, and $u=(k_1-p_2)^2$.
 In addition, $\epsilon^{\mu}(k_1,\lambda_1)$ will denote the $\gamma$  
polarization vector.
 We have evaluated the amplitudes using dimensional
regularization, with the help of the program 
FeynCalc~\cite{Mertig:1990an} and the numerical libraries
FF~\cite{vanOldenborgh:1990yc,vanOldenborgh:1990wn}.

 The result for the total amplitude of the reaction
$e \gamma \to e h^0 (H^0)$ can be written as:
\begin{equation}
 {\cal M}={\cal M}_{\gamma}+{\cal M}_Z+{\cal M}^{\rm box}_Z
 +{\cal M}^{\rm box}_W,
\end{equation}
where
${\cal M}_\gamma$, ${\cal M}_Z$, ${\cal M}^{\rm box}_Z$, and 
${\cal M}^{\rm box}_W$
correspond to the sets of diagrams (1), (2), (3), and (4), respectively.
 The amplitudes coming from the
$\phi^0 \gamma^* \gamma$ and $\phi^0Z^* \gamma$ couplings can be written
as follows:     
\begin{equation}
{\cal M}_{\gamma,Z}
 =
 \frac{i\alpha^2m_W}{4s^3_wc^4_w}
 \bar{u}(p_2)\gamma^\nu(a_{\gamma,Z}-b_{\gamma,Z}\gamma_5)
 u(p_1)\epsilon^{\mu}(k_1,\lambda_1)F_{\gamma,Z}
 (k_1\cdot k_2g_{\mu\nu}-k_{2\mu}k_{1\nu}),
\end{equation}  
where   
\begin{equation}
 F_{\gamma}=\frac{4s^2_wc^4_w}{m^2_Wt}
 \left(
  \sum_f 2f_{\phi}N_c Q^2_f F^{1/2}_f + v_{\phi}F^1_{\gamma}+s_{\phi}F^0
 \right),
\end{equation}
\begin{equation}
 F_Z=\frac{c^4_w}{m^2_W(m^2_Z-t)}
  \left(
   -\sum_f\frac{f_{\phi}C^f_VN_cQ_f}{c^2_w}F^{1/2}_f
   +v_{\phi}F^1_Z-\frac{s_{\phi}c_{2w}}{2c^2_w}F^0
  \right),
\end{equation}
with $a_{\gamma}=1$, $b_{\gamma}=0$, $a_Z=1-4s^2_w$, $b_Z=1$, 
$t_w = \frac{s_w}{c_w}$, and $c_{2w}=c^2_w-s^2_w$. 
The coefficients $f_{\phi}$, $v_{\phi}$, and $s_{\phi}$ characterize the
$\phi^0 \bar{f}f$, $\phi^0 ZZ$, and $\phi^0 H^{\pm}H^{\mp}$ couplings
in the MSSM, and are given by:
\begin{displaymath}
 f_\phi=
 \left\{
  \begin{array}{ll}
   \frac{\sin \alpha}{\sin \beta}, & \textrm{$H^0 \bar{u}u$}\\
   \frac{\cos \alpha}{\cos \beta}, & \textrm{$H^0 \bar{d}d$}\\
   \frac{\cos \alpha}{\sin \beta}, & \textrm{$h^0 \bar{u}u$}\\
  -\frac{\sin \alpha}{\cos \beta}, & \textrm{$h^0 \bar{d}d$},
  \end{array}
 \right.
\end{displaymath}
\begin{displaymath}
 v_\phi=  
 \left\{
 \begin{array}{ll}
  \cos(\beta-\alpha), & \textrm{$\phi^0=H^0$}\\
  \sin(\beta-\alpha), & \textrm{$\phi^0=h^0$},
 \end{array}
 \right.
\end{displaymath}
\begin{displaymath}
 s_\phi=
  \left\{
  \begin{array}{ll}
   \cos(\beta-\alpha)-\frac{1}{2c^2_w}\cos2\beta \cos(\beta+\alpha), &
   \textrm{$\phi^0=H^0$}\\
   \sin(\beta-\alpha)+\frac{1}{2c^2_w}\cos2\beta \sin(\beta+\alpha), &
   \textrm{$\phi^0=h^0$}.
  \end{array}
 \right.
\end{displaymath}
The functions $F^{1/2}_f$, $F^0$, $F^1_{\gamma}$ and $F^1_Z$ arising from 
fermion, scalar and gauge boson loops are given in the Appendix B.

 The amplitude for the contributions of the $Z$-mediated box diagram,
including the related  $\phi^0e^{*-}e^-$ triangle graphs, is the following:
\begin{eqnarray}
 {\cal M}^{\rm box}_Z   
 &=&
 \frac{i\alpha^2m_W}{4s^3_wc^4_w}\bar{u}(p_2)\gamma^{\nu}
  \left(a_Z-\gamma_5\right)^2u(p_1)\epsilon^\mu(k_1,\lambda_1) \nonumber
\\
 & 
 \times
 &
 \left[-A(t,s,u)\left(k_1\cdot p_1g_{\mu\nu}-p_{1\mu}k_{1\nu}\right)
 +A(t,u,s)(k_1\cdot p_2g_{\mu\nu}-p_{2\mu}k_{1\nu})\right],
\end{eqnarray}   
where the functions $A(t,s,u)$ and $A(t,u,s)$ are presented in the Appendix B.

 The amplitude for the $W$-mediated box diagrams and its related triangle
graphs is given by:
\begin{eqnarray}
 {\cal M}^{\rm box}_W
 &=&
 \frac{i\alpha^2m_W}{4s^3_wc^4_w}
 \bar{u}(p_2)\gamma^{\nu}(1-\gamma_5)^2u(p_1)\epsilon^{\mu}(k_1,\lambda_1)
 \left\{
  A_{12}(t,s,u)
 \right.
 \nonumber \\
  &  
  \times
  &
  \left.
  (k_1\cdot p_1g_{\mu\nu}-p_{1\mu}k_{1\nu})
  -A_{21}(t,s,u)(k_1\cdot p_2 g_{\mu\nu}-p_{2\mu}k_{1\nu})
 \right\},
\end{eqnarray}
where the functions $A_i$ and $A_{ij}$ are given in the Appendix B.

 After squaring the total amplitude, the corresponding cross-section
for this process  is given by:
\begin{equation}
 \hat{\sigma}
 =
 \frac{1}{16\pi s^2}\int_{m^2_{\phi}-s}^{0}dt
 \arrowvert \overline{\cal M} \arrowvert^2,
\end{equation}
where
\begin{equation}
 \arrowvert \overline{\cal M}\arrowvert^2
 =  
 \frac{\alpha^4m^2_W}{64s^6_wc^8_w}(-t)
 \left[
  \left(s^2+u^2\right)F_{\gamma Z}+s^2F_s+u^2F_u
 \right],
\end{equation}
with
\begin{equation}
 F_{\gamma Z}
 =
 \arrowvert F_\gamma \arrowvert^2 + a\arrowvert
 F_Z \arrowvert^2+2a_ZRe\left(F_ZF^*_{\gamma}\right),
\end{equation}
\begin{eqnarray}
 F_s
 &=&
 \arrowvert A_{12}(t,s,u)\arrowvert^2
 +c\arrowvert 
 A(t,s,u)\arrowvert^2 +2Re
 \left\{
  [A_{12}(t,s,u)-2aA(t,s,u)]F^*_{\gamma}
 \right.
 \nonumber \\
  &+&
 \left.
  \left[
   4bA_{12}(t,s,u)-2dA\left(t,s,u\right)
  \right]F^*_Z-2A_{12}\left(t,s,u\right)A^*\left(t,s,u\right)
 \right\},
\end{eqnarray}
\begin{equation}
 F_u=F_s(s \leftrightarrow u),
\end{equation}
where we have defined $a=1+a^2_z$, $b=1+a_z$, $c=1+6a^2_z+a^4_z$,
and $d=a_z(a^2_z+3)$.

 Finally, in order to obtain the total cross-section $(\sigma_T)$, one
needs to convolute  $\hat{\sigma}$ with the photon distribution, namely:
\begin{equation}
 \sigma_T
 =
 \frac{1}{S}\int_{m^2_{\phi}}^{0.83S}f_{\gamma}(\frac{s}{S})\hat{\sigma}(s)ds,
\end{equation}
 where $S$ denotes the squared c.m. energy of the $e^+e^-$-system, and the
photon distribution  is given by:
\begin{equation}
 f_{\gamma}
 =
 \frac{1}{D(\xi)}
 \left[ 
  1-x + \frac{1}{1-x} - \frac{4x}{\xi(1-x)}
  + \frac{4x^2}{\xi^2(1-x)^2}
 \right],
\end{equation}
 where
\begin{equation}
 D(\xi)
 =
 \left(
  1-\frac{4}{\xi}-\frac{8}{\xi^2}
 \right)
  \log(1+\xi)
  +\frac{1}{2}+\frac{8}{\xi}-\frac{1}{2(1+\xi)^2}.
\end{equation}
Notice that, as in Ref.~\cite{Gabrielli:1997ix}, one should use, instead
of the photon distribution, an exactly monochromatic initial photon
beam which can help in distinguish the phisical effects related to the
particular collision process from details depending on the final
realization of the laser beam. 

 To evaluate the cross-section for $h^0$ and $H^0$,
we have taken $m_t= 175\;$ GeV, and the values for the electroweak
parameters given in the table of particle properties~\cite{Caso:1998tx}.
 In fact, as it was discussed in Ref.~\cite{Gabrielli:1997ix},  
the contributions of the boxes can be neglected.
 The cross section for $h^0$ is about 4.2 fb for $m_{A^0}>200$
GeV and $E_{c.m.}=500$ GeV, as can be seen from Fig. 2.
 With the expected luminosity at the future linear
colliders~\cite{Kuhlman:1996rc,Accomando:1997wt,Tauchi:1998ty} 
of about 50 fb$^{-1}$/yr, it will be possible to observe up to about 210
$h^0 + e$ events, which should allow to study the properties of the Higgs
boson. 
 Among them, it will be interesting to test
its spin and CP-even nature by studying the angular distributions,
however this aspect is beyond the goal of the present work, where
we are mainly interested in determining if the MSSM Higgs sector
can be tested at electron-photon colliders.
 Total cross-sections for the production of the
heavy Higgs boson $H^0$ as a function of the pseudoscalar mass 
$m_{A^0}$ and for different values of $\tan\beta$
are shown in Fig. 3.
 These graphs present the total cross-section as a function of the
pseudoscalar mass ($m_{A^0}$) for several values of  $\tan\beta$ for c.m.
energy of 500 GeV.

\subsection{Production of $A^0$}
 On the other hand the amplitude for the production of the
pseudoscalar contains only contributions from the fermionic
triangles. Thus, the amplitude takes the form:
\begin{equation}
 {\cal M} = {\cal M}_\gamma +{\cal M}_Z,
\end{equation}
where
\begin{equation}
{\cal M}_\gamma
 =
 \frac{- i \alpha^2 Q_f^2 N_C f_\beta m_W}{s_w}
 \bar{u}(p_2)\gamma^\nu u(p_1) \epsilon^\mu (k_1,\lambda_1)
 \frac{4 m_f^2}{m_W^2} C_0 \left(t, m_A^2, m_f^2 \right)
 \left(
  \frac{\epsilon_{\mu\nu\alpha\beta}k_1^\alpha k_2^\beta}{t}
 \right), 
\end{equation}  
and
\begin{equation}
{\cal M}_Z
 =
 \frac{i \alpha^2 Q_f C_V^f N_C f_\beta m_W}{4 s^3_w c^2_w}
 \bar{u}(p_2) \gamma^\nu 
 \left( 
  C_V^e - C_A^e \gamma_5
 \right)
 u(p_1)\epsilon^\mu(k_1,\lambda_1)
 \frac{4 m_f^2}{m_W^2} C_0 \left(t, m_A^2, m_f^2 \right)
 \left(
  \frac{\epsilon_{\mu\nu\alpha\beta}k_1^\alpha k_2^\beta}{t - m_Z^2}
 \right),
\end{equation}  
with $C_0 \left(t, m_A^2, m_f^2 \right)=
C_0 \left(t, m_A^2, 0, m_f^2, m_f^2, m_f^2 \right)$ the Passarino-Veltman 
three-point scalar function written in the notation of the FeynCalc program 
and 
\begin{equation}
f_\beta=\left\{
\begin{array}{ll}
\cot\beta,&\textrm{$f=u$}\\
\tan\beta,&\textrm{$f=d$},
\end{array}
\right.
\end{equation}
The total squared amplitude is
\begin{eqnarray}
 \arrowvert \overline{\cal M}\arrowvert^2
 &
 =
 &
 \frac{\alpha^4 Q_f^4 N_C^2 f_\beta^2}{s^2_w}
 \frac{-t}{m_A} \frac{s^2 + u^2}{t^2}
 \left[
  1- 
  \frac{C_V^e C_V^f}{2 s_w^2 c_w^2 Q_f} 
  \frac{t}{t - m_Z^2}
 \right.
  \\
  &  
  +
  & 
 \left. 
  {C_V^f}^2 
  \frac{{C_V^e}^2 + {C_V^f}^2}{16 s_w^4 c_w^4 Q_f^2}
  \left(\frac{t}{t - m_Z^2}\right)^2
 \right]
 \arrowvert
  2 m_f^2 C_0\left(t, m_A^2, m_f^2 \right)
 |^2.
\end{eqnarray}
 To obtain the total cross-section we use the expressions
written previously for the CP-even Higgs bosons (Eqs. 13-15).
 Results are shown in Fig. 4, where we plot again the total cross-section
as a function of the pseudoscalar mass ($m_{A^0}$),
for several values of  $\tan\beta$ and for c.m.
energy of 500 GeV.   It can be noted that the
cross-section for $A^0$ is smaller than the one
resulting for $h^0/H^0$.  Thus, in this case it will be
more difficult to detect the signal.

\subsection{Backgrounds and exclusion contours}

 The final signature for the reaction $e\gamma \to e \phi^0_i$
depends on the decay of the Higgs boson.
 For $h^0$, the dominant mode is into $b\bar b$, whereas for $H^0$ and
$A^0$ this decay can also be relevant for some regions of parameter space.
 To evaluate the B.R. of the Higgs bosons we shall assume that the decays
into SUSY modes (i.e. charginos, neutralinos, sfermions) are not allowed
and take the relevant equation for the decay widths from~\cite{kane:90}.
Our results are in agreement with the ones obtained in the literature,
for instance in ref. \cite{Spira:1995rr}.
Thus, we shall concentrate on the signature coming from the decay
$\phi^0_i \to b\bar b$. In this case the main background comes from
$e\gamma \to e b\bar{b}$, which receives contribution from 8
graphs at tree level; we have evaluated numerically this processes using
CompHep~\cite{Boos:1994xb,Baikov:1997zr}.
 Following Dicus et al.~\cite{Dicus:1996in}, we have also imposed a cut on
the angular distribution of the outgoing electron 
($|\cos\theta| < 0.98$, relative to the incident photon) that reduces
significantly the background, while retaining most of the signal
rate.
 To determine the region of parameter space, which is taken
as the plane $\tan\beta$-$m_{A^0}$, where the Higgs signal is detectable,
we have proceeded as follows:
for each value of $m_{A^0}$ we evaluate the Higgs masses and the 
cross-sections,then we find the value for $\tan\beta$ where the cross-section
of the signal is above the background at the 3 and $5\sigma$ level.

 In Fig. 5, we show the regions in the plane 
$\tan\beta$-$m_{A^0}$ where the cross-section coming
from $h^0$ has detectable values.
 The region to the right from the heavy line is where the 
signal is detectable, above backgrounds at the 3$\sigma$ level.
The dashed line denotes the contour at the 5$\sigma$ level.
 It can be seen that the signal is above the
backgrounds for a significant region of parameter space. 

 On the other hand, Fig. 6 shows the region where the cross-section
from $H^0$ reaches detectable values.  Again, 
 The region to the left from the heavy (dashed) line is where the 
signal is detectable at the 3 (5) $\sigma$ level.
 It can be noticed that this region covers the sector of parameter space
where it is more difficult to detect the light $h^0$.
 Thus, the cross-sections for $h^0/A^0$ and $H^0/A^0$ play a 
complementary role in providing a detectable signal for 
the full parameter space.
 Moreover, we also notice from the superposition of Figs. 5 and 6,
that there is a small region where the two Higgs boson signals can
be detected, which will allow to distinguish clearly between the
MSSM and the SM Higgs sectors.

 Finally, we also want to stress the fact that this reaction offers
a unique opportunity to obtain a clean measurement of the Higgs mass,
thanks to the possibility to measure the outgoing electron momentum.
The mass of the Higgs boson is related to the maximal energy of
the electron as:
\begin{equation}
 m_{h^0} = \sqrt{s} - E_{max}.
\end{equation}
 Thus, the precision attained for the electron energy will translate into
a good determination of the Higgs mass.

 On the other hand, the largest values of the cross-section for $A^0$
are obtained for large values of $\tan\beta$ (which is usually assumed to be
at most of order 50); however, we find that
even in this case the resulting cross-section does not seem to give a
detectable signal. Moreover, for large values of $\tan\beta$
there appears a mass-degeneracy between  $A^0$ and $h^0$ or $H^0$, which
will make difficult to distinguish the individual signals. 
In this case, it will be neccesary to optimize the cuts to be able
to detect the pseudoscalar $A^0$, and to separate it from the largest
signals coming from $h^0$ and $H^0$. Otherwise,
one would have to add the respective signals; for instance, one could
imposed the criteria that whenever the difference between the
Higgs masses is less than  5 GeV (a conservative criteria given
the possibility to obtain a precise measurement of the Higgs mass),
the individual contributions to the signal must be combined. We find
that this only helps to enlarge the exclusion contours for
$\tan\beta$ above 30.

Another option to detect $A^0$ (and a heavy $H^0$ too) is to use the
planned second and third stages of the future e-gamma
colliders~\cite{Kuhlman:1996rc,Accomando:1997wt,Tauchi:1998ty}, which
could reach an energy of 1 TeV and 2 TeV, with integrated luminosities
that could reach 125 and 500 $fb^{-1}$, respectively.

\section{The charged Higgs}
 Now, turning to the production of the charged Higgs, we observe that
it can also proceed through the 1-loop reaction:
\begin{eqnarray}
 \gamma(k_1)+e^-(p_1) &\to& H^-(k_2)+\nu_e(p_2).
\end{eqnarray}

 The diagrams encountered in the calculation of the
$e^- \gamma \rightarrow \nu H^- $ are shown in Fig.~\ref{diagrch}.
 They include: triangle graphs with bosons and fermions in the 
loop (Fig.~\ref{diagrch}a), 
seagull-type graphs with bosonic contributions (Fig.~\ref{diagrch}b-c),
and finally those with self-energy 
insertions~(Fig.~\ref{diagrch}d, \ref{diagrch}e, \ref{diagrch}f).
 Fig.~\ref{diagrch}f gives a vanishing contribution for 
massless external fermions.

\subsection{Production of $H^\pm$}
 Our result for the total amplitude is written as:

\begin{eqnarray}
 {\cal M} 
 &
 =
 & 
 \frac{i \alpha^2}{2 \sqrt{2} {\rm s}^3_w m_W (t - m^2_W)} 
 \bar{u}(p_2) 
 \gamma^\nu
 (1-\gamma_5)
 u(p_1) \epsilon^\mu (k_1,\lambda_1)
 \nonumber
 \\
 &
 \times
 &
 \left[
  \left(
   V_f + V_{H^+ \phi^0} + V_{W \phi^0}
  \right)
  \left( 
   k_{2\mu} k_{1\nu} -g_{\mu \nu}k_1 \cdot k_2 
  \right)
  + i A_f \epsilon_{\mu \nu \alpha \beta} k^\alpha_1 k^\beta_2 
 \right]
\end{eqnarray}
 where $V_f$, $V_{H^+ \phi^0}$, $V_{W \phi^0}$ and $A_f$ denote the 
contribution from the different sets of graphs shown in 
Fig.~\ref{diagrch} and are given in the Appendix C.

 Fig.~\ref{plotchsig} shows the cross-section for
our process for several values of $\tan\beta$ (= 1.5, 5, 10).
 The cross section is small, about 0.5 fb for 
$m_{H^\pm} = 200$ GeV and $\rm E_{c.m.}=500$ GeV,
which can give 25 events
with an expected future linear colliders luminosities of
50~fb$^{-1}$/yr.
 In Fig.~\ref{plotchall} we compare the cross section from 
$e \gamma \rightarrow H^- \nu$, with the pair production 
$e^+ e^- \rightarrow H^+ H^-$,
$e \gamma \rightarrow H^+ H^- e$
and $\gamma \gamma \rightarrow H^+ H^-$.
 The production mechanisms 
$e^+ e^- \rightarrow H^+ H^-$,
$\gamma \gamma \rightarrow H^+ H^-$,
have been discussed in the literature\cite{Bowser-Chao:1993ji}.
 On the other hand, the reaction 
$e \gamma \rightarrow H^+ H^- e$ is evaluated using the 
Williams-Weizsacker approximation (for the second photon)
and we use the sub-reaction $\gamma \gamma \rightarrow H^+ H^-$.
 It can be seen that the single production dominates only for 
large Higgs masses i.e. for  values that lay beyond the threshold
for pair production.

\subsection{Backgrounds}

 In this paper we only show (Fig.~\ref{plotchsig}) the results for fixed
values of $\tan\beta= 1.5,5,10$, and as can be appreciated it turns out
that the resulting cross-section has smaller results,
which hardly seem detectable.
 In order to determine if the signal is detectable, one needs to 
consider the potential 
backgrounds, which depend on the Higgs mass, since this
determines the decay signatures.
 For instance, if $m_{H^\pm} < m_t + m_b$, the dominant decay 
mode is $H^+ \rightarrow \tau \nu$ which reaches a branching ratio
of order 1 for most values of $\tan \beta$.
 In this case the background will come from the production 
$\gamma e \rightarrow W^* \nu$, which for masses 
$m_{H^\pm} \approx m_W$ will be much larger than the signal
(it reaches $\sigma \simeq 4 \rm pb$)
and probably will not allow detection.
 For masses somehow larger, both the signal and the
background will be more suppressed and 
the question of detectability will depend on the experimental 
ability to identify the decay mode and to reconstruct the charged 
Higgs mass.

 For heavier Higgs masses, $M_{H^\pm} > m_t + m_b$, the dominant 
Higgs decay is into $t+\bar{b}$, and in this case
one needs to compare the signal with the background arising 
from single top production,
$e \gamma \rightarrow \nu \bar{t} b$ for $M_{H^\pm} > m_t + m_b$,
which according to the results of Boos et 
al.~\cite{Boos:1997rd}, has
$\sigma \simeq 15 \rm fb$ for $\sqrt{s} = 500$ GeV and 
$m_t = 175$GeV.
 Thus we can see that at the level of total cross-section the 
background is again larger than the signal. 
 However if one makes a cut in the invariant mass of the $t-b$
system, then it will be possible to reduce the background.

\section{Conclusions}
 We have studied the production of the neutral CP-even and charged
Higgs boson of the MSSM ($h^0,H^0, H^\pm$) at future $e \gamma$ colliders,
through the reactions $e\gamma \rightarrow e h^0, eH^0, e A^0, \nu H^-$. 
 The amplitudes are evaluated using a non-linear $R_{\xi}$-gauge, 
which greatly simplifies the calculation.
 The resulting cross section indicates that it 
is possible to detect the light neutral Higgs boson ($h^0$) 
for most values of parameters. 
 On the other hand, detection of the heavy neutral
Higgs bosons $H^0$ seems possible only for light values of $m_{A^0}$.
 We have determined the regions in the plane $\tan\beta$-$m_{A^0}$ where
the signal is above the backgrounds.
 The cross-sections for $h^0$ and $H^0$ play a complementary role, since
the region where  $H^0$ reaches detectable values occurs precisely in 
the region where it is more difficult to detect
the light $h^0$.
 Thus, both reactions allow to cover the full plane 
$\tan\beta-m_{A^0}$ with at least one detectable signal.
However, it is found that the possibility to distinguish the MSSM
from the SM case, through detection of both $h^0$ and $H^0$ signals,
occurs only for a limited region of parameter space.

 On the other hand, the results for the pseudoscalar and the
charged Higgs boson have smaller values of the
cross-section, which seems difficult to detect. 

\acknowledgments 
We acknowledge financial support from CONACYT and SNI (M\'exico).
U.C. whish to acknowledge useful discussions with Borut Bajc and Goran
Senjanovi\'c.

\appendix 
\section{The non-linear gauge}
 In the evaluation of the above processes, we have used
a Feynman-t'Hooft version of the nonlinear 
gauge~\cite{Fujikawa:1972fe,Bace:1975qi,Gavela:1981ri,Deshpande:1982mi,%
Hernandez:1999xn}
that greatly simplify the calculation. 
 In order to appreciate the advantages that offer a nonlinear
gauge, we find convenient to compare it with the conventional
linear gauge for the SM. 
 It is well known that in a linear gauge only the charged fermion
sector presents explicit electromagnetic gauge symmetry ($\rm U_{em}(1)$),   
since the gauge fixing procedure used for the SU(2) sector
destroys manifest $\rm U_{em}(1)$ symmetry in
the charged gauge boson ($W^\pm$) and ghosts sectors ($c,\bar{c}$),
when they are considered separately,
because the gauge functional used to define the $W$-propagator does 
not transform covariantly under the $\rm U_{em}(1)$ symmetry.
 It follows that instead of obeying naive Ward identities, these sectors 
are related through Ward-Slavnov identities. 
 In order to obtain a finite and $\rm U_{em}(1)$-invariant
result for higher-order (loop) calculations, one must sum over
the contributions arising from the $W$ gauge boson, 
the charged ghosts, and the $W^\pm G^\mp$ combined effects,  
$G^\pm$ denotes the would-be Goldstone boson. 
 On the other hand, the functional used to define the 
$W$-propagator in the non-linear gauge contains the electromagnetic
covariant derivative.  
 Thus, the $\rm U_{em}(1)$ symmetry is respected by
each charged sector of the SM.
 It follows that a finite and $\rm U_{em}(1)$ gauge invariant result is 
obtained for each type of diagram containing a given kind of charged 
particles. 
 Moreover, the number of diagrams involved is considerably
reduced because there are no $W^\pm G^\mp$ combined effects. 
 In the SM, the functionals that define the $W$, $Z$, and 
$\gamma$ propagators are given by:
\begin{eqnarray}
 f^+ & = & \bar{D}_\mu W^{+\mu}-i\xi m_WG^+, \nonumber \\
 f^Z & = & \partial_\mu Z^\mu - \xi m_Z G^0,  \\
 f^A & = & \partial_\mu A^\mu,             \nonumber 
\end{eqnarray}
where $D_\mu = \partial_\mu - ig'B_\mu$, with
$B_\mu = -s_w Z_\mu + c_w A_\mu $, $G^0$ is the neutral would-be Goldstone.
 $\xi$ is the gauge parameter, which in general is different for
each of the three gauge bosons. However, 
in the non-linear Feynman-t'Hooft version of this gauge, 
one takes $\xi=1$ for all sectors. 
 The gauge fixing Lagrangian is given by:
\begin{equation}
 {\cal L}_{GF}
 =
 -\frac{1}{\xi}f^+f^--\frac{1}{2\xi}(f^Z)^2-\frac{1}{2\xi}(f^A)^2,
\end{equation}
 which in turn removes the $W^\pm G^\mp \gamma$ and $W^\pm G^\mp Z$
vertices from the Higgs kinetic energy term. 
 Notice that the $f^+f^-$ term
is $\rm U_{em}(1)$ gauge invariant, which implies that the Fadeev-Popov 
${\cal L}_{FP}$ Lagrangian is also $\rm U_{em}(1)$-invariant. 
 It is also possible to proceed further with this scheme and remove
more un-physical vertices, such as   
$\phi^0 W^\pm G^\mp \gamma(Z)$ and $G^0 W^\pm G^\mp \gamma(Z)$. 
 This procedure requires to define the functionals $f^\pm$
nonlinearly both in the vector and scalar parts.
 However, for the  present purposes it is sufficient to use the
scheme presented above.

 In this case only the couplings in the gauge and ghost sectors are
modified. We shall give the lagrangian for these sectors in a form that
clarifies the role of the electromagnetic gauge invariance.
For this, we define the derivative:
$\hat{D}_\mu= \partial_\mu -i g W^3_\mu$,
with $W^3_\mu= c_w Z_\mu + s_w A_\mu$. This operator shares with the
derivative $D$ defined above the property of containing the e.m.
covariant derivative. After adding the  gauge-fixing lagrangian,
one obtains the couplings for the gauge bosons ($W,Z,A$),
which are contained in:

\begin{eqnarray}
{\cal L}&=&-\frac{1}{2}( \hat{D}_\mu W^+_\nu-\hat{D}_\nu W^+_\mu)^\dagger
                      ( \hat{D}^\mu W^{+\nu}-\hat{D}^\nu W^{+\mu})
       - ig (s_w F_{\mu\nu}+c_w Z_{\mu\nu}) W^{-\mu}W^{+\nu} \nonumber \\
      & & - ig 
     \left[
      s_w F_{\mu\nu}+c_w Z_{\mu\nu} + i\frac{g}{2}
       \left( 
        W^-_\mu W^+_\nu - W^-_\nu W^+_\mu
       \right) 
     \right]
     W^{-\mu}W^{+\nu} \nonumber \\
   & & 
   -\frac{1}{\xi}\left(
     \bar{D}_\mu W^{+\mu}
    \right)
    \left(
     \bar{D}_\nu W^{+\nu}
    \right)^\dagger
    + m^2_W  W^-_\mu W^{+\mu} \nonumber \\
   & & 
   - \frac{1}{4} Z_{\mu\nu} Z^{\mu\nu} 
   + \frac{1}{2} m^2_Z Z_\mu Z^\mu 
   - \frac{1}{2\xi} 
   \left(
    \partial_\mu Z^\mu
   \right)^2  
   - \frac{1}{4}  F_{\mu\nu} F_{\mu\nu}
   - \frac{1}{2\xi} 
   \left(\partial_\mu A^\mu\right)^2
\end{eqnarray}
where 
$F_{\mu\nu}=\partial_\mu A_\nu-\partial_\nu A_\mu$,
$Z_{\mu\nu}=\partial_\mu Z_\nu-\partial_\nu Z_\mu$.
 Notice that, apart from the gauge fixing term $\partial_\mu A^\mu$,
this lagrangian is $\rm U_{em}(1)$ invariant.

 On the other hand, the ghost sector is substantially modified
in the non-linear gauge; however, since the gauge-fixing functionals
are U(1) invariant, the charged part of this sector is also invariant.
 We present now the corresponding Fadeev-Popov lagrangian, written in such
a way that $\rm U_{em}(1)$ invariance is explicit, namely:
\begin{eqnarray}
{\cal L}_{FPG}
 &=&
 -\bar{c}^-
 \left[ 
  \bar{D}_\mu \hat{D}^\mu + \xi m_W
  \left(
   m_W + \phi^0 + iG^0
  \right)
 \right] 
 c^+ \nonumber \\
 & & 
 -igc_W \bar{c}^-\bar{D}_\mu
 \left(
  c_Z W^{+\mu}
 \right)
 -ie \bar{c}^-\bar{D}_\mu(c_{\gamma} W^{+\mu}) \nonumber\\
 & & 
 -i\frac{gs^2_W}{c_W} W^{+\mu}\bar{c}^-
 \left(
  \partial_\mu c_Z
 \right)
 +ie W^{+\mu}\bar{c}^-
 \left(
  \partial_\mu c_\gamma
 \right) \nonumber \\
 & & 
 -igc_W W^{+\mu}
 \left(
  \partial_\mu \bar{c}_Z
 \right) 
 c^- -ie W^{+\mu} 
 \left(
  \partial_\mu \bar{c}_\gamma
 \right) c^- \nonumber \\
 & & 
 -g c_{2w} \xi m_Z G^+ 
 \left( 
  \bar{c}^- c_Z +\bar{c}_Z c^- 
 \right)
  -e \xi m_W G^+ 
 \left(
  \bar{c}^- c_\gamma +\bar{c}_\gamma c^- 
 \right) + h.c. \nonumber \\
 & & 
 -\bar{c}_Z 
 \left[ \Box + \xi m_Z 
  \left(
   m_Z + \phi^0
  \right) 
 \right] 
 c_Z - \bar{c}_\gamma \Box c_\gamma 
\end{eqnarray}
 where $c^\pm (\bar{c}^\pm), c_Z (\bar{c}_Z), c_\gamma (\bar{c}_\gamma)$ 
denote the pairs of ghosts associated with the
$W,Z,A$ gauge bosons respectively.
 The phase-convention for the charged ghost is $(c^+)^\dagger=c^-$ 
and $(\bar{c})^\dagger=\bar{c}^-$.
 In addition, one has:
$\phi^0=\cos(\alpha-\beta) H^0+\sin(\alpha-\beta) h^0$, where
$h^0,H^0$ denote the
light and heavy neutral Higgs bosons of the MSSM. Notice that
neither the charged Higgs ($H^\pm$) nor the CP-odd Higgs
bosons ($A^0$) appear in this lagrangian, which can be understood
if one reminds that these fields do not appear in the definition of
the gauge fixing terms, and also because they do not receive a v.e.v..

Finally, we would like to mention the properties of the couplings
of gauge bosons with the higgs and goldstone bosons. First, in this
gauge the couplings $W^+ G^- \gamma$ and $W^+ G^-Z$ are absent by construction, whereas
the couplings $G^+ G^- Z (\gamma), H^+H^- Z (\gamma), \phi^0 W^+W^-$ and
$W^+H^- \phi^0$ do not depend on the gauge-fixing procedure.
 However, the couplings $\phi^0 G^+ G^-$ or $\phi^0 c^+c^-$ can depend on
the choice of the gauge-fixing terms, when both the vector and
scalars sectors are chosen non-linearly.

 In conclusion, the Feynman rules needed for our calculation, which 
are depending of the gauge fixing procedure, arise from the 
lagrangians of Eqs. (A3-A4) written above.

\section{Loop functions associated to neutral scalars}
\label{a2}
In this appendix we present the various loop functions coming from the 
$e^-\gamma \to e^-\phi^0$ processes. The functions arising from fermion, 
scalar and gauge boson loops, which characterize the $\phi^0 \gamma^* \gamma$ 
and $\phi^0 Z^*\gamma$ couplings, are given by:

\begin{eqnarray}
 F^{1/2}_f
 &
 =
 &
 \frac{4m^2_f}{m^2_{\phi}-t}
  \left\{
   1-\frac{1}{2}
   \left(
    m^2_{\phi}-t-4m^2_f
   \right)C_0\left(t,m^2_{\phi},m^2_f\right) 
  \right. 
   \nonumber 
  \\
  &
  +
  &
  \left.
  \frac{t}{m^2_{\phi}-t}
  \left[
   B_0\left(m^2_\phi,m^2_f\right)-B_0\left(t,m^2_f\right)
  \right]
 \right\},
\end{eqnarray}
\begin{equation}
 F^0=-\frac{4m^2_W}{m^2_{\phi}-t}
 \left\{
  1+2m^2_{H^{\pm}}C_0\left(t,m^2_{\phi},m^2_{H^\pm}\right)
  +\frac{t}{m^2_{\phi}-t}\left[B_0\left(m^2_{\phi},m^2_{H^\pm}\right)
  -B_0\left(t,m^2_{H^\pm}\right)\right]
 \right\},
\end{equation}
\begin{eqnarray}
 F^1_{\gamma} 
 &=&
 -\frac{4m^2_W}{m^2_{\phi}-t}
 \left\{
  3+\frac{m^2_{\phi}}{2m^2_W}+
  \left[
   3\left(1-\frac{m^2_{\phi}}{2m^2_W}\right)+\frac{2t}{m^2_W}
  \right]
  2m^2_WC_0\left(t,m^2_{\phi},m^2_W\right) 
 \right.
 \nonumber \\
  &+&
 \left. 
 \left(3+\frac{m^2_\phi}{2m^2_W}\right)\left(\frac{t}{m^2_\phi-t}\right)
  \left[
   B_0
    \left(
     m^2_{\phi},m^2_W
    \right)-B_0
    \left(t,m^2_W\right)
  \right]
 \right\},
\end{eqnarray}
\begin{eqnarray}
 F^1_Z
 &=&
 -4\left(3-t^2_w\right)m^2_WC_0\left(t,m^2_{\phi},m^2_W\right)
 +\frac{2m^2_W}{m^2_\phi-t}
 \left[
  5+\frac{m^2_\phi}{2m^2_W}
  -\left(1+\frac{m^2_{\phi}}{2m^2_W}\right)t^2_w
 \right]
  \nonumber \\
  &
  \times
  &
  \left\{
  1 +
  2m^2_WC_0(t,m^2_{\phi},m^2_W)
  +\frac{t}{m^2_\phi-t}
  \left[
   B_0\left(m^2_{\phi},m^2_W\right)-B_0\left(t,m^2_W\right)
  \right]
 \right\},
\end{eqnarray}
where 
$C_0(t,m^2_{\phi},m^2_i)=C_0(0,t,m^2_{\phi},m^2_i,m^2_i,m^2_i)$,
$B_0(m^2_{\phi},m^2_i)=B_0(m^2_{\phi},m^2_i,m^2_i)$ and
$B_0(t,m^2_i)=B_0(t,m^2_i,m^2_i)$ are Passarino-Veltman three- and two-point
scalar functions written in the notation of the FeynCalc program.

The loop functions coming from $Z$-mediated box diagram, including the related 
$\phi^0 e^{*-}e^-$ triangle graphs, are given by:

\begin{eqnarray}
 A(t,s,u)
 &=&
 \frac{1}{2st}     
 \left\{
  \frac{s-m^2_Z}{s}\left[m^2_Z(s+u)-su\right]D_0(1,2,3,4)
  +(s-m^2_Z)\left[C_0(1,2,4) 
 \right.
 \right.
 \nonumber \\
  &+&
 \left.
  \frac{u}{s}C_0(1,2,3)-\frac{t+u}{s}C_0(2,3,4)
  +\frac{1}{s}\left(t-s-\frac{2m^2_Z st}{(s+t)(s-m^2_Z)}\right)
  C_0(1,3,4)\right]
 \nonumber \\
  &+& 
 \left.
  \frac{2t}{s+t}
  \left[
   B_0(3,4)-B_0(1,3)
  \right]
 \right\}.
\end{eqnarray}  
$A(t,u,s)$ is obtained from $A(t,s,u)$ by means of the interchange
$s \leftrightarrow u$.
 The arguments of the scalar functions are given by:
\begin{eqnarray}
 D_0(1,2,3,4)   
 &=&
 D_0\left(0,s,m^2_{\phi},u,0,0,m^2_e,m^2_e,m^2_Z,m^2_Z\right),\nonumber \\
 C_0(1,2,4)
 &=& C_0\left(0,s,0,m^2_e,m^2_e,m^2_Z\right), \nonumber \\
 C_0(1,2,3)
 &=& C_0\left(0,0,u,m^2_e,m^2_e,m^2_Z\right), \nonumber \\
 C_0(2,3,4)
 &=&
 C_0\left(s,0,m^2_\phi,m^2_Z,m^2_e,m^2_Z\right), \nonumber \\
 C_0(1,3,4)
 &=&
 C_0\left(0,m^2_\phi,u,m^2_e,m^2_Z,m^2_Z\right), \nonumber \\
 B_0(3,4)
 &=&
 B_0\left(m^2_\phi,m^2_Z,m^2_Z\right), \nonumber \\
 B_0(1,3)
 &=&
 B_0\left(u,m^2_e,m^2_Z\right).
\end{eqnarray}

Finally, the loop functions associated to the $W$-mediated box 
diagrams and its related triangle graphs are the following:

\begin{eqnarray}
 A_{12}(t,s,u) &=& 2c^4_w\left[A_1(t,s,u)+A_2(t,u,s)\right], \nonumber \\
 A_{21}(t,s,u) &=& 2c^4_w\left[A_2(t,s,u)+A_1(t,u,s)\right],
\end{eqnarray}
with
\begin{eqnarray}
 A_1(t,s,u)
 &=&
 \frac{1}{2st}
 \left\{
  \frac{s-m^2_W}{s}\left[m^2_W(s+u)+st\right]D_0(1,2,3,4)
 \right. \nonumber \\
  &+&
  (s-m^2_W)\left[C_0(2,3,4)
  -\frac{t}{s}C_0(1,3,4)-\frac{s+u}{s}C_0(1,2,3)
  \right.
 \nonumber \\
  &+&
 \left.
  \left.
  \frac{u+t}{s}C_0(1,2,4)\right]
  +\frac{2t}{s+u}\left[B_0(1,2)-B_0(1,3)\right]
 \right\},
\end{eqnarray}  
\begin{eqnarray}
 A_2(t,s,u)
 &=&
 \frac{1}{2tu}
 \left\{
  \left[
   \frac{t+u-m^2_W}{u}\left(m^2_W(s+u)-st\right)-2m^2_Wt
  \right]
  D_0(1,2,3,4) 
 \right.
  \nonumber \\
  &+&
  \left(t+u-m^2_W\right)
  \left[
   \frac{s}{u}C_0(2,3,4)
   +\frac{t}{u}C_0(1,3,4)
  \right.
   \nonumber \\
   &-&
   \left.
   \frac{s+u}{u}C_0(1,2,3)+\frac{u^2-2ut-t^2}{u(t+u)}C_0(1,2,4)
  \right] 
  \nonumber \\
  &+&
 \left. 
  \frac{2t}{t+u}\left[B_0(2,4)-B_0(1,2)\right]
  +\frac{2t}{s+u}\left[B_0(1,3)-B_0(1,2)\right]
 \right\}.
\end{eqnarray}
The scalar functions have the following arguments:
\begin{eqnarray}
 D_0(1,2,3,4)   
 &=&
 D_0\left(0,0,0,m^2_\phi,t,s,m^2_W,0,m^2_W,m^2_W\right),\nonumber \\
 C_0(2,3,4)
 &=&
 C_0\left(0,0,s,0,m^2_W,m^2_W\right), \nonumber \\
 C_0(1,3,4)
 &=& C_0\left(0,0,t,m^2_W,0,m^2_W\right), \nonumber \\
 C_0(1,2,3)
 &=&
 C_0\left(t,0,m^2_\phi,m^2_W,m^2_W,m^2_W\right), \nonumber \\
 C_0(1,2,4)
 &=&
 C_0\left(0,s,m^2_\phi,m^2_W,0,m^2_W\right), \nonumber \\
 B_0(1,2)
 &=&
 B_0\left(m^2_{\phi},m^2_W,m^2_W\right), \nonumber \\
 B_0(1,3)
 &=&
 B_0\left(t,m^2_W,m^2_W\right), \nonumber \\
 B_0(2,4)
 &=&
 B_0\left(s,0,m^2_W\right).
\end{eqnarray}
 The expressions for $A_1(t,u,s)$ and $A_2(t,u,s)$ can be obtained from
the respective $A_1(t,s,u)$ and $A_2(t,s,u)$ through the interchange
$s \leftrightarrow u$.
\section{Loop functions associated to charged scalars} 
 In this appendix we shall write the explicit formulae for the 
functions $V_f$, $V_{H^\pm \phi^0}$, $V_{W \phi^0}$ and $A_f$ 
using a notation that, hopefully, facilitates its use by the 
interested reader.
 Our convention for the scalar functions $B_0$, $C_0$ is the same 
as in the previous Appendix.
In this case, the contributions
arising from loops with fermions, scalars and gauge bosons,
have the following expression:

\begin{eqnarray}
 V_f
 &
 =
 &
 \sum_f
 \frac{N_c}{t - m_{H^\pm}^2}
 \left\{
  \left(
   Q_u + Q_d 
  \right)
  \left(
   m_d^2 \tan\beta + m_u^2 \cot\beta
  \right)
 \right. 
  \nonumber
  \\
  &+&
 \left.
  2Q_d
  \left[
   m_d^2 
   \left(
    m_d^2 \tan\beta + m_u^2 \cot\beta
   \right)
   -
   \left(
    t - m_{H^\pm}^2
   \right)
   m_d^2 \tan\beta
  \right]
  C_0
  \left(
   0, m^2_{H^\pm}, t, m_d^2, m_d^2, m_u^2
  \right)
 \right. 
  \nonumber
  \\
  &+&
 \left.
  2Q_u
  \left[
   m_u^2 
   \left(
    m_d^2 \tan\beta + m_u^2 \cot\beta
   \right)
   -
   \left(
    t - m_{H^\pm}^2
   \right)
   m_u^2 \cot\beta
  \right]
  C_0
  \left(
   0, m^2_{H^\pm}, t, m_u^2, m_u^2, m_d^2
  \right)
 \right. 
  \nonumber
  \\
  &+&
 \left.
  \left[
   \frac{m_d^2 \tan\beta + m_u^2 \cot\beta}{t - m_{H^\pm}^2} 
   \left(
    m_u^2 - m_d^2
    +
    \left(
     Q_u + Q_d 
    \right) 
    m_{H^\pm}^2   
   \right)
 \right.
  \right. 
   \nonumber
   \\
   &+&
 \left.   
  \left.
   2
   \left( 
    Q_d m_d^2 \tan\beta + Q_u m_u^2 \cot\beta
   \right)
  \right]   
  \left[
   B_0
   \left(
    t, m_d^2, m_u^2
   \right)
   -
   B_0
   \left(
    m_{H^\pm}^2, m_d^2, m_u^2
   \right)   
  \right]
 \right. 
  \nonumber
  \\
  &+&
 \left.
  m_u m_d
  \frac{m_u^2 - m_d^2}{m_{H^\pm}^2}
  \left[
   B_0
   \left(
    m_{H^\pm}^2, m_d^2, m_u^2
   \right)
   -
   B_0
   \left(
    0, m_d^2, m_u^2
   \right)   
  \right]
 \right\}, 
\end{eqnarray}
and
\begin{eqnarray}
 A_f
 &
 =
 &
 \sum_f
 N_c
 \left\{
  Q_d m_d^2 \tan\beta
  C_0
  \left(
   0, m^2_{H^\pm}, t, m_d^2, m_d^2, m_u^2
  \right)
  -
  Q_u m_u^2 \cot\beta 
  C_0
  \left(
   0, m^2_{H^\pm}, t, m_u^2, m_u^2, m_d^2
  \right)
 \right. 
  \nonumber  
  \\
  &+&
 \left.
  \frac{m_d^2 \tan\beta + m_u^2 \cot\beta}{t - m_{H^\pm}^2}
  \left[
   B_0
   \left(
    t, m_d^2, m_u^2
   \right)
   -
   B_0
   \left(
    m_{H^\pm}^2, m_d^2, m_u^2
   \right)   
  \right]
 \right\},
\end{eqnarray}
\begin{eqnarray}
 V_{H^+ \phi^0} 
 &
 =
 & 
  \sum_{\phi^0 = h^0, H^0}
  \left(
   -v_{\phi^0} s_{\phi^0}
  \right)
  \frac{m_W^2}{t - m^2_{H^\pm}}
  \left\{
   1 
   + 
   2m^2_{H^\pm}C_0
   \left(
    0, t, m^2_{H^\pm}, m^2_{H^\pm}, m^2_{H^\pm}, m^2_{\phi^0}
   \right) 
  \right. \nonumber 
  \\
  &+&
  \left.
   \left(
    \frac{m^2_{\phi^0} - m^2_{H^\pm} - t}{t - m^2_{H^\pm}}
   \right)
   \left[
    B_0
    \left(
     t, m^2_{H^\pm}, m^2_{\phi^0}
    \right)
    -
    B_0
    \left(
     m^2_{H^\pm}, m^2_{H^\pm}, m^2_{\phi^0}
    \right)    
   \right]
  \right. \nonumber 
  \\
  &+&
  \left.
   \left(
    \frac{m^2_{\phi^0} - m^2_{H^\pm}}{m^2_{H^\pm}}
   \right)
   \left[
    B_0
    \left(
     0, m^2_{H^\pm}, m^2_{\phi^0}
    \right)
    -
    B_0
    \left(
     m^2_{H^\pm}, m^2_{H^\pm}, m^2_{\phi^0}
    \right)
   \right]
  \right\},
\end{eqnarray}
\begin{eqnarray}
 V_{W \phi^0} 
 &
 =
 & 
 \sum_{\phi^0 = h^0, H^0} 
 \left(
  -\sigma_{\phi_0} v_{\phi_0}
 \right)
 \frac{m_W^2}{t - m^2_{H^\pm}}
 \left\{
  \frac{m^2_{H^\pm} - m^2_{\phi^0} + m^2_W}{2 m^2_W} 
  \right. \nonumber 
  \\
  &+&
  \left.
  \left(
   3 m^2_{H^\pm} - m^2_{\phi^0} + m^2_W - 2t
   \right) 
   C_0
   \left(
    0, m^2_{H^\pm}, t, m^2_W, m^2_W, m^2_{\phi^0}
   \right) 
  \right. \nonumber 
  \\
  &+&
  \left.
   \frac{\left(
          m^2_{H^\pm} - m^2_{\phi^0} - m^2_W
         \right)
         \left(
          m^2_{\phi^0_i} - m^2_W -t
         \right)
         -4 m^2_W
         \left(
          m^2_{H^\pm} -t
         \right)}%
         {2 m^2_W
         \left(
          t - m^2_{H^\pm} 
         \right)}
  \right. \nonumber 
  \\
  &\times&
  \left.
   \left[
    B_0
    \left(
     t, m^2_W, m^2_W
    \right)
    -
    B_0
    \left(
     m^2_{H^\pm}, m^2_W, m^2_W
    \right)    
   \right]
  \right. \nonumber 
  \\
  &+&
  \left.
   \frac{\left(
          m^2_{\phi^0} - m^2_W
         \right)
         \left(
          m^2_{H^\pm} - m^2_{\phi^0} + m^2_W
         \right)
         }{2 m^2_W m^2_{H^\pm}}
   \left[
    B_0
    \left(
     0, m^2_W, m^2_{\phi^0}
    \right)
    -
    B_0
    \left(
     m^2_{H^\pm}, m^2_W, m^2_{\phi^0}
    \right)
   \right]
  \right\},
\end{eqnarray}
 where
$\sigma_{H^0} = -v_{h^0}$,
$\sigma_{h^0} = -v_{H^0}$
and
$\phi^0_1 = H^0$, 
$\phi^0_2 = h^0$.


\providecommand{\href}[2]{#2}\begingroup\raggedright\endgroup

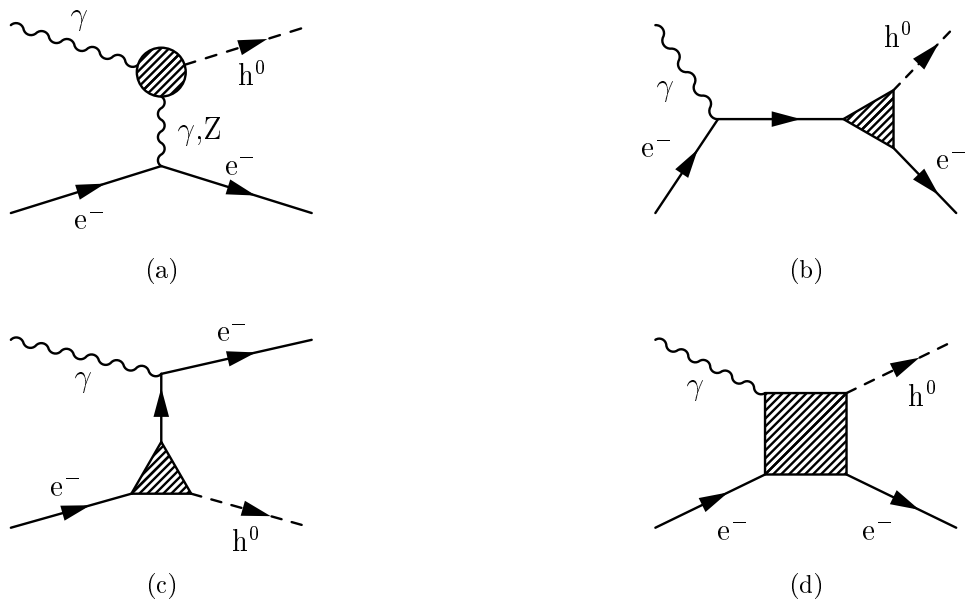
\begin{figure}
\caption{Classification of graphs that contributes to the 
          reaction $e \gamma \rightarrow e h^0$.%
          }
 \label{diagrnh}
\end{figure}

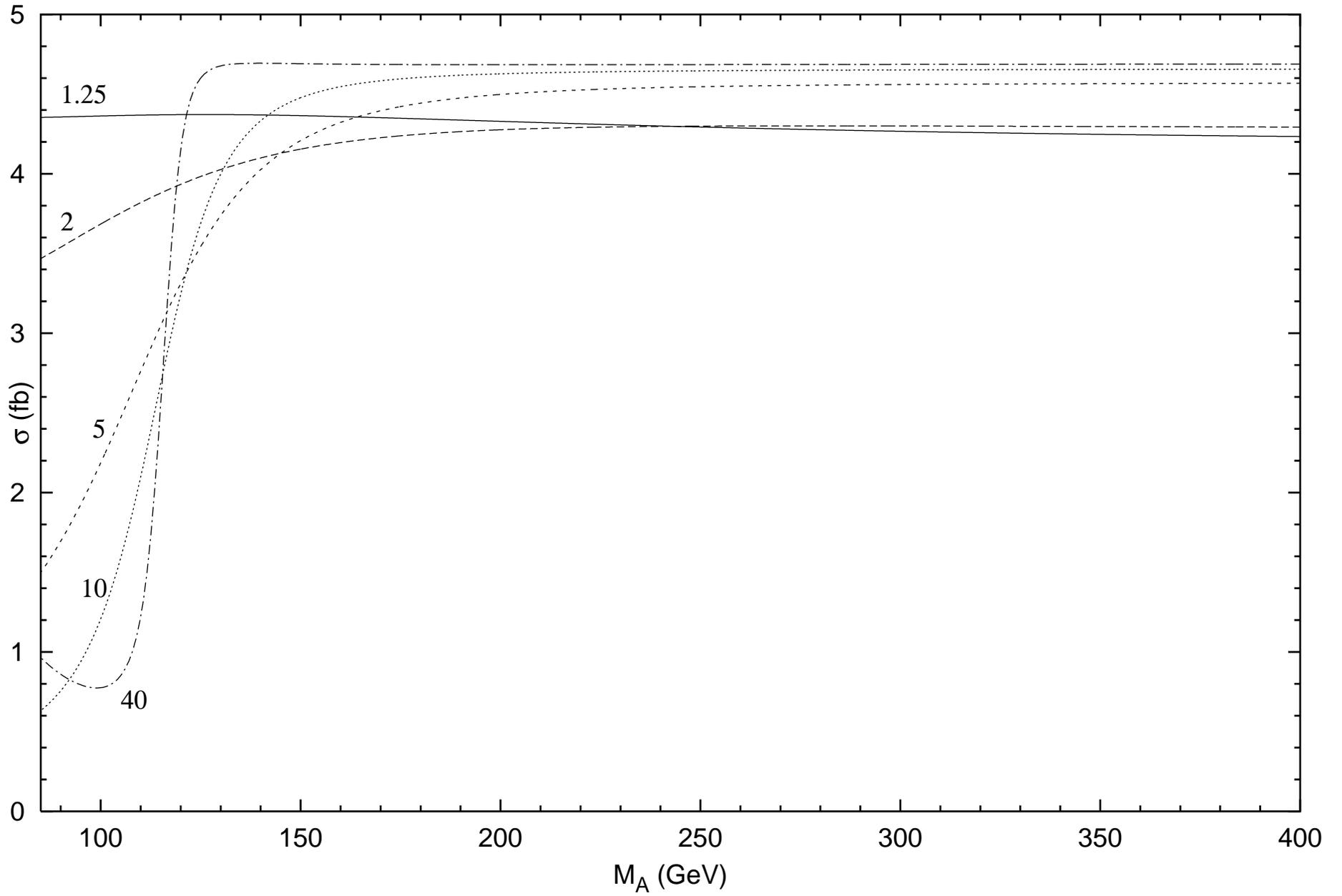
\begin{figure}
 \caption{Total cross section for the reaction 
          $e \gamma \rightarrow e h^0$ 
          for different values of $\tan\beta$, 
          with $\protect\sqrt{s}$ = 500 GeV.
           }
 \label{plotlhsig}
\end{figure}

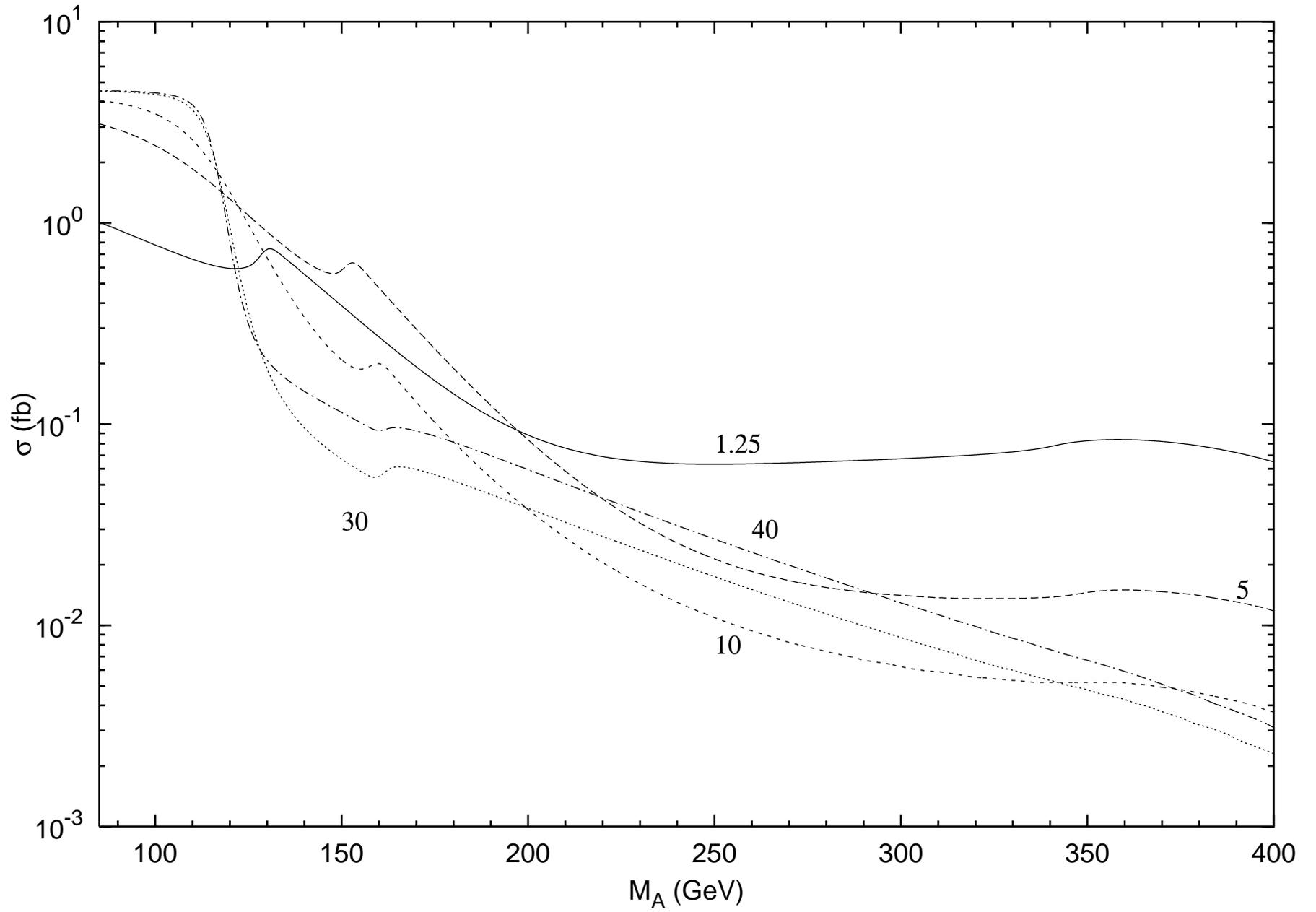
\begin{figure}
 \caption{Total cross section for the reaction 
          $e \gamma \rightarrow e H^0$ 
          for different values of $\tan\beta$, 
          with $\protect\sqrt{s}$ = 500 GeV.
           }
 \label{plothhsig}
\end{figure}

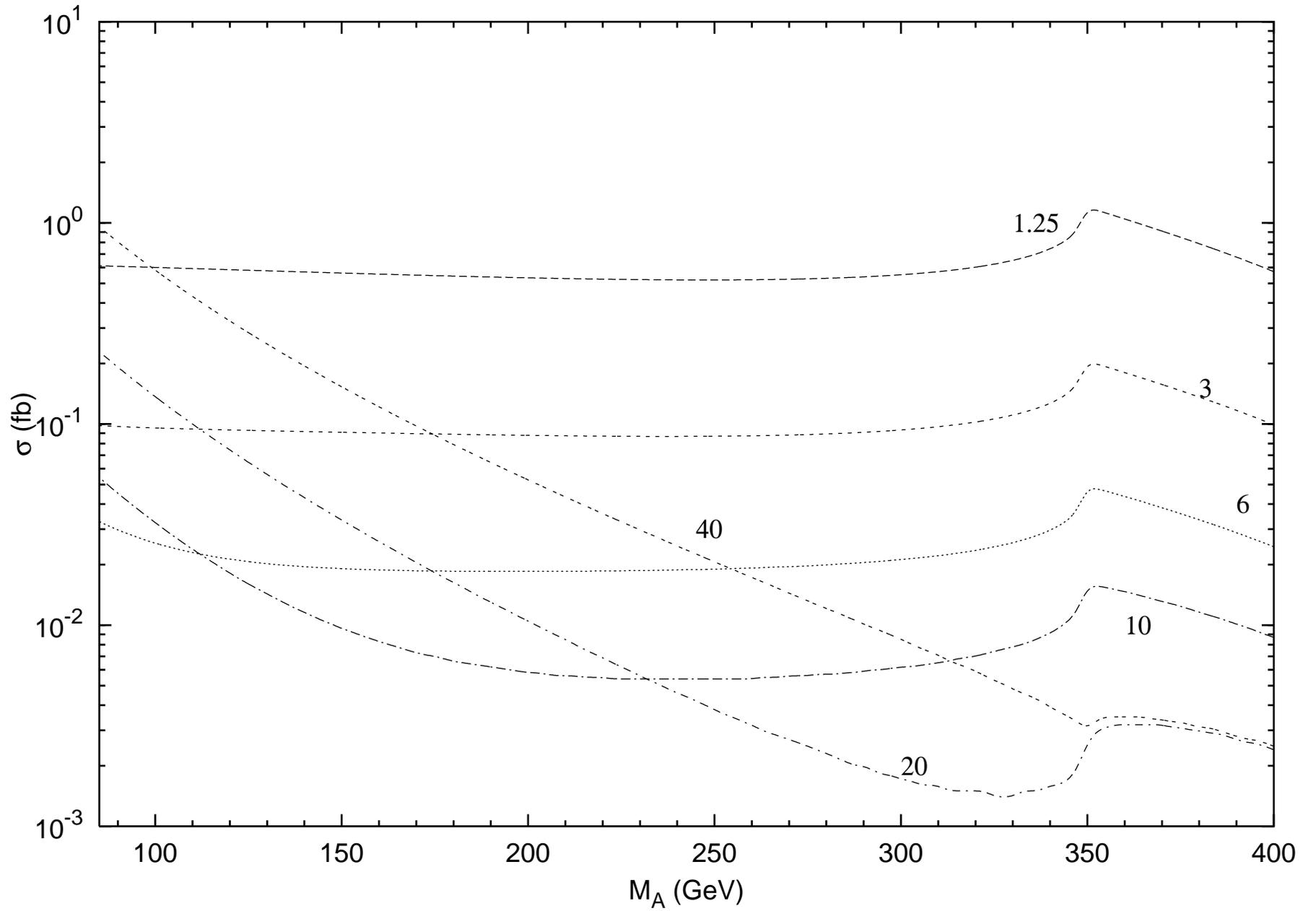
\begin{figure}
 \caption{Total cross section for the reaction 
          $e \gamma \rightarrow e A^0$ 
          for different values of $\tan\beta$, 
          with $\protect\sqrt{s}$ = 500 GeV GeV.
           }
 \label{plotapsig}
\end{figure}

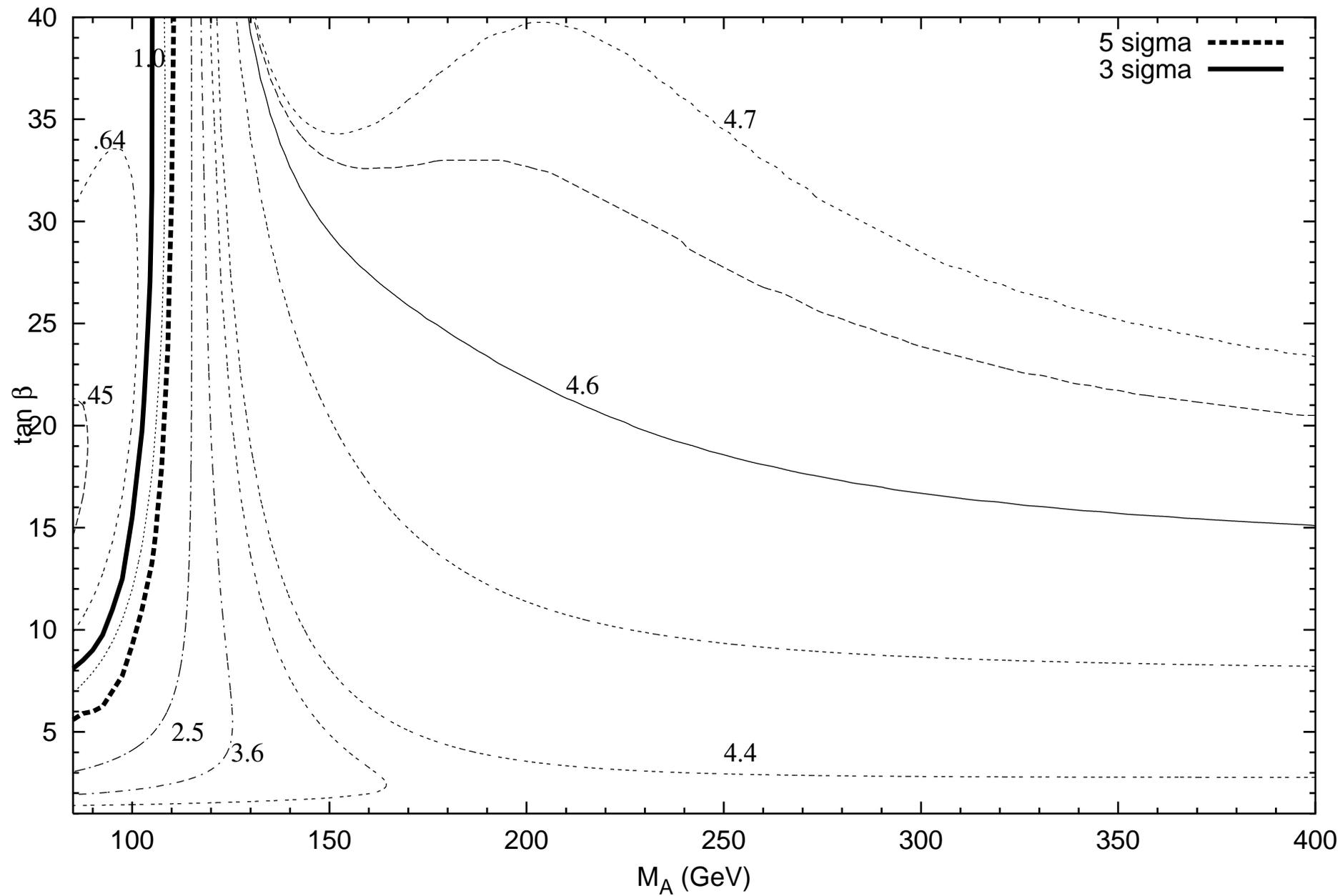
\begin{figure}
 \caption{Cross section contour lines for the reaction 
          $e \gamma \rightarrow e h^0$, 
          for different values of $\sigma$(fb), 
          with $\protect\sqrt{s}$ = 500 GeV.
           The region to the right from the heavy (dashed) line is where
          the 
          signal is detectable (above backgrounds at the 3 (5) $\sigma$
          level).
          }
 \label{plotlhtb}
\end{figure}

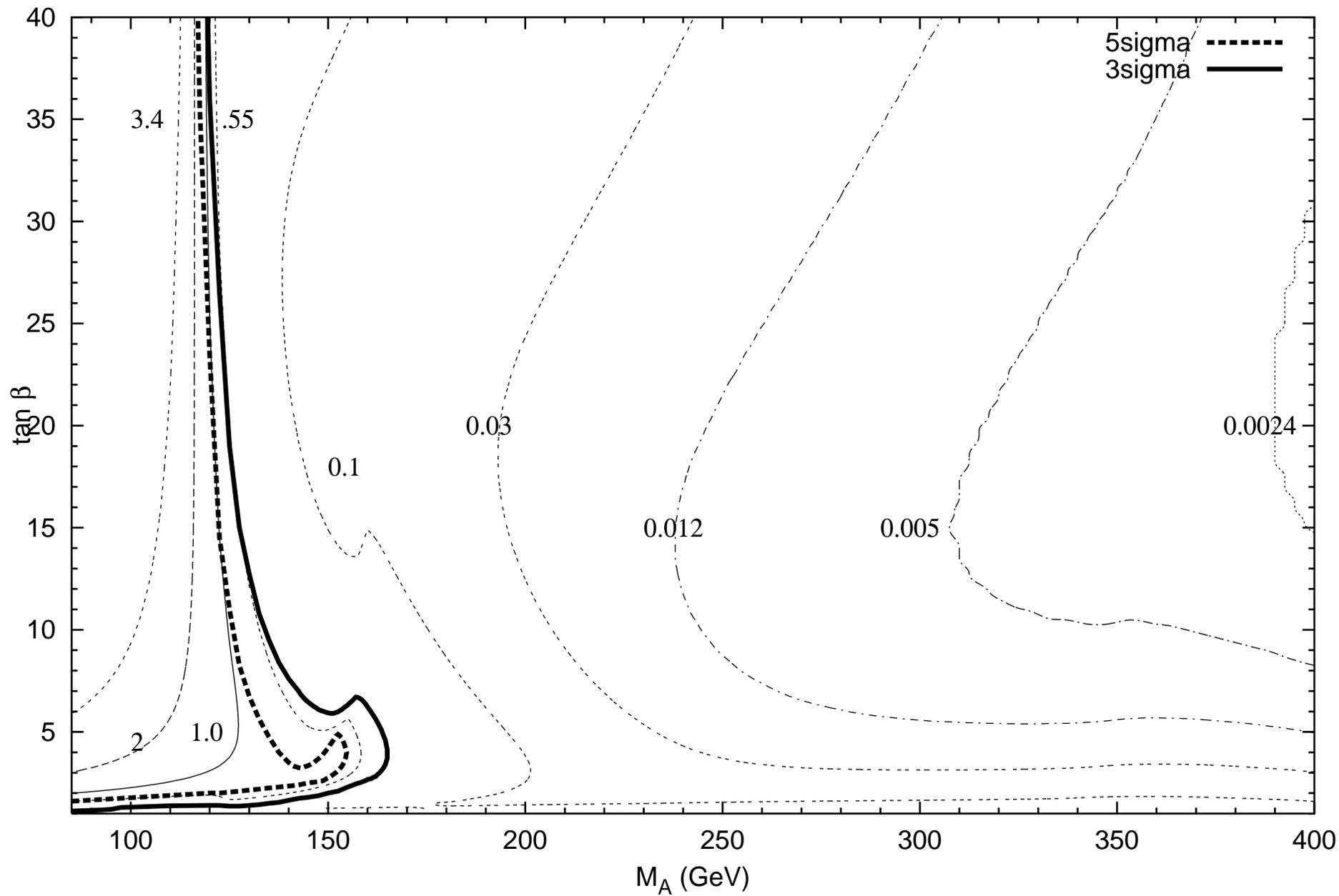
\begin{figure}
 \caption{Cross section contour lines for the reaction 
          $e \gamma \rightarrow e H^0$, 
          for different values of $\sigma$(fb), 
          with $\protect\sqrt{s}$ = 500 GeV.
           The region to the left from the heavy (dashed) line is where
          the 
          signal is detectable (above backgrounds at the 3 (5)$\sigma$
level).}
 \label{plothhtb}
\end{figure}

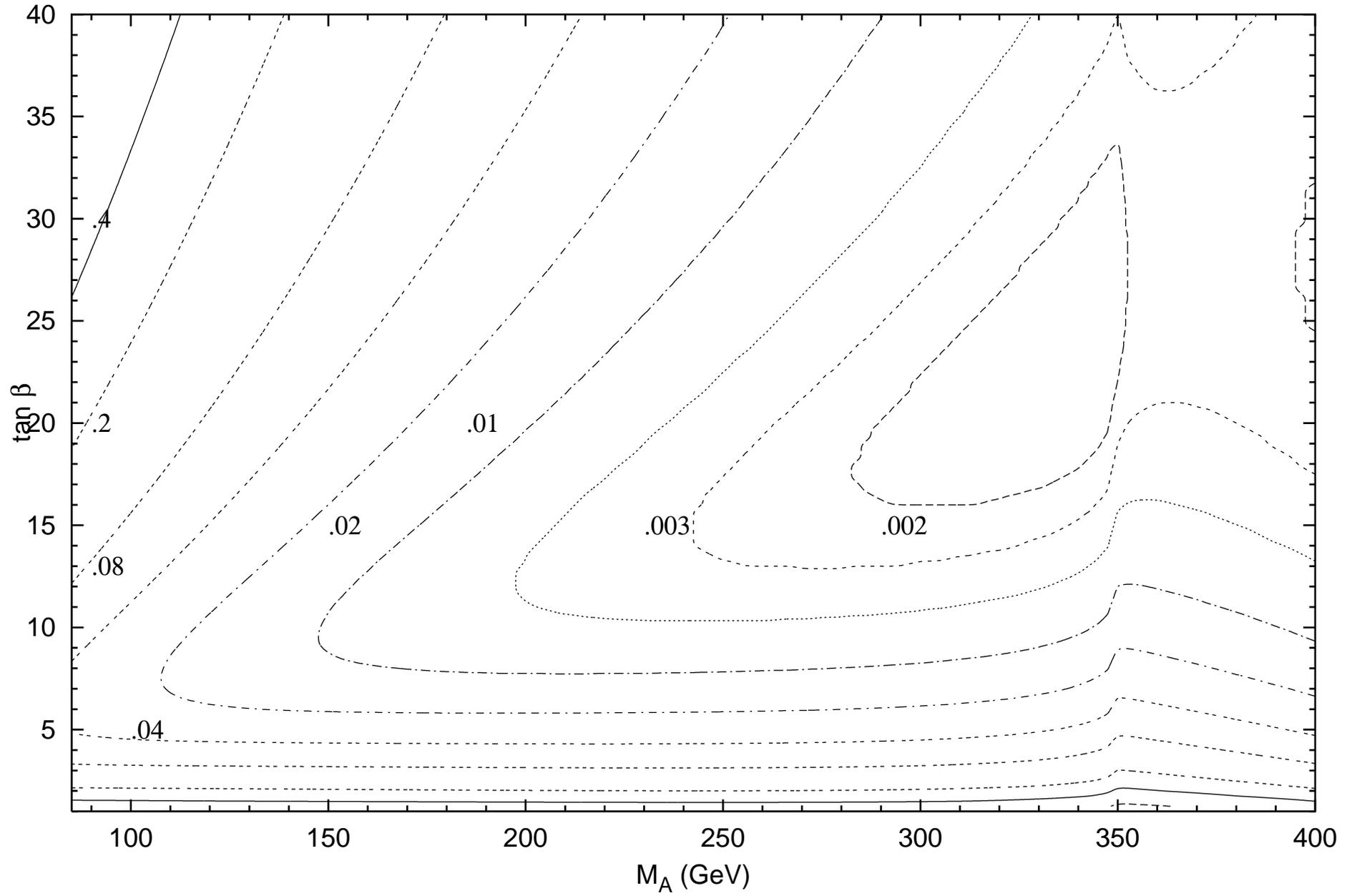
\begin{figure}
 \caption{Cross section contour lines for the reaction 
          $e \gamma \rightarrow e A^0$, 
          for different values of $\sigma$(fb), 
          with $\protect\sqrt{s}$ = 500 GeV.
          }
 \label{plotaptb}
\end{figure}

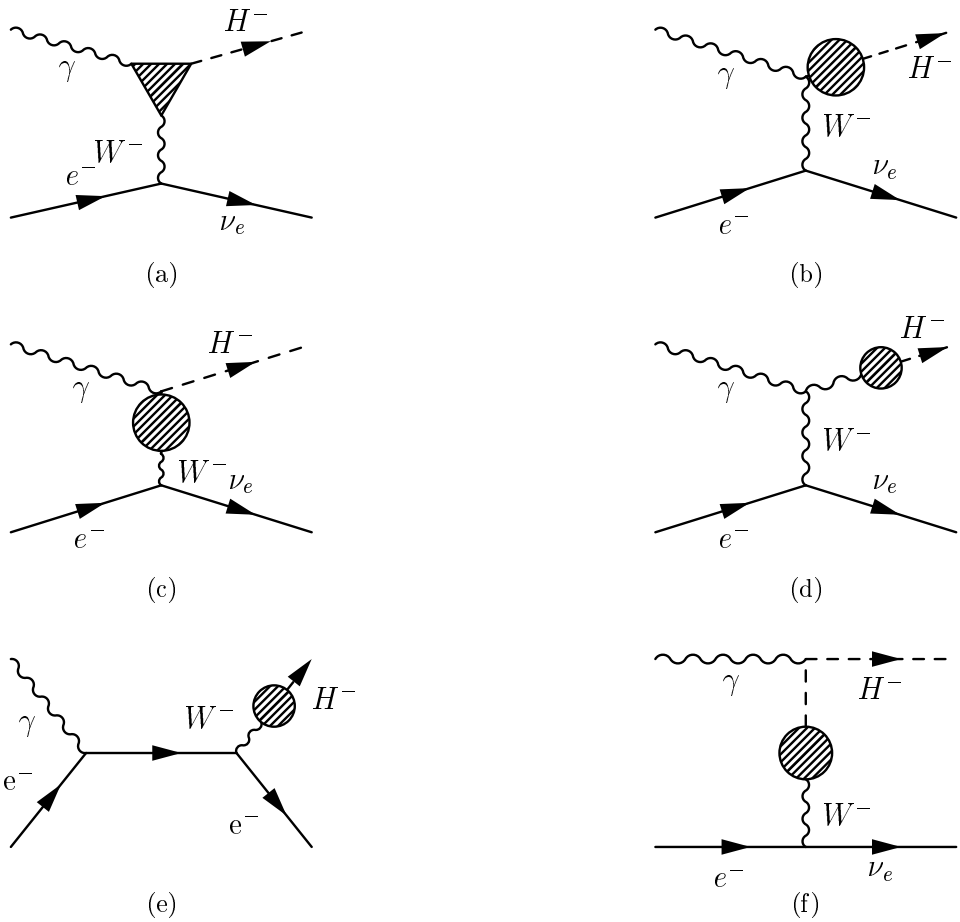
\begin{figure}
\caption{Classification of graphs that contributes to the 
          reaction $e^- \gamma \rightarrow \nu_e H^-$.%
          }
 \label{diagrch}
\end{figure}

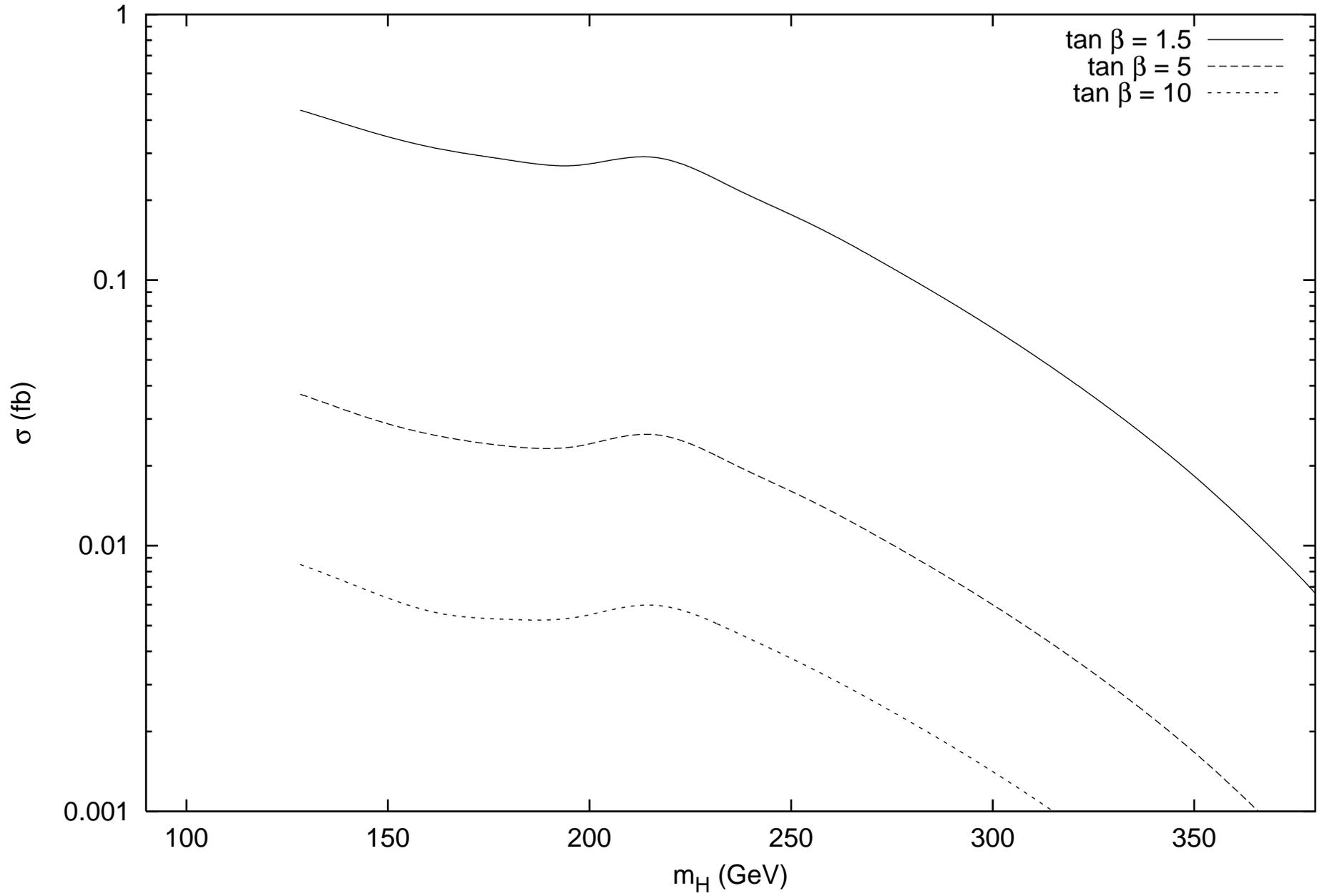
\begin{figure}
 \caption{Cross section contour lines for the reaction 
          $e^- \gamma \rightarrow \nu H^-$, 
          for different values of $\tan \beta$, 
          with $\protect\sqrt{s}$ = 500 GeV.}
 \label{plotchsig}
\end{figure}

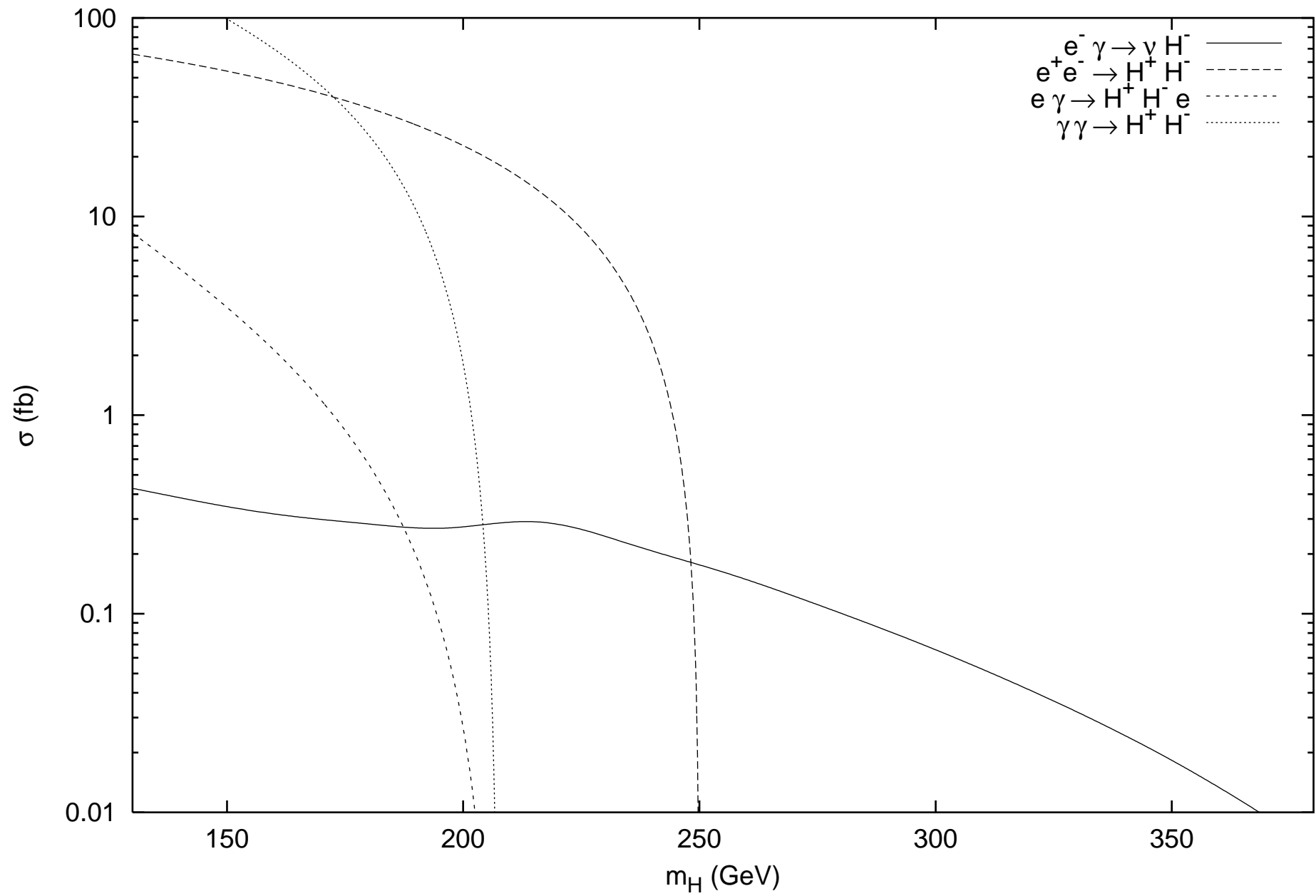
\begin{figure}
 \caption{Cross section for the reaction 
          $e^- \gamma \rightarrow \nu H^-$, with 
          $\protect\sqrt{s}$ = 500 GeV compared with different production
          channels. }
 \label{plotchall}
\end{figure}

\end{document}